\documentclass[prb,preprint]{revtex4-1}

\usepackage{amsmath} 
\usepackage{amsfonts} 
\usepackage{graphicx} 

\begin{document}

\title{Berry phase and the unconventional quantum Hall effect in graphene}

\author{Jiamin Xue}
\email{xue@email.arizona.edu}

\affiliation{Microelectronic Research Center,  The University of Texas at Austin, Austin, Texas 78758, USA}

\date{\today}

\begin{abstract}
The Berry phase of $\pi$ in graphene is derived in a pedagogical way. The ambiguity of how to calculate this value properly is clarified. Its connection with the unconventional quantum Hall effect in graphene is discussed.
\end{abstract}

\maketitle

\section{Introduction}

Since its first isolation \cite{2004paper} in 2004, graphene has attracted great research interest. It is not only a magic material for realizing many research ideas \cite{the rise of graphene}, but also an elegant model system for teaching and studying many concepts in condensed matter physics. An enormous amount of literature on this subject is available. If beginners of graphene want to understand some concepts, they often need to track down a long list of references. For the case of Berry phase in graphene, this can be a confusing experience. Despite its importance \cite{2005 geim paper, 2005 kim paper}, there has been some ambiguity of how to calculate it (see, e.g. the influential review paper in Ref. \onlinecite{rev of modern physics} and the debate about the Berry phase in bilayer graphene in Ref. \onlinecite{winding number}). The aim of this paper is to derive the Berry phase in graphene in a pedagogical way, clarify the ambiguity, and reveal its connection with the unconventional quantum Hall effect.

This paper is organized as follows. In Section \ref{band} the basic band structure of graphene is briefly introduced. In Section \ref{berry} the $\pi$ Berry phase in graphene is derived. In Section \ref{quantumhall} how the $\pi$ Berry phase manifests itself in the unconventional quantum Hall effect is discussed. 

\section{Band structure of graphene}
\label{band}

\begin{figure}[h!]
\centering
\includegraphics[width=3in]{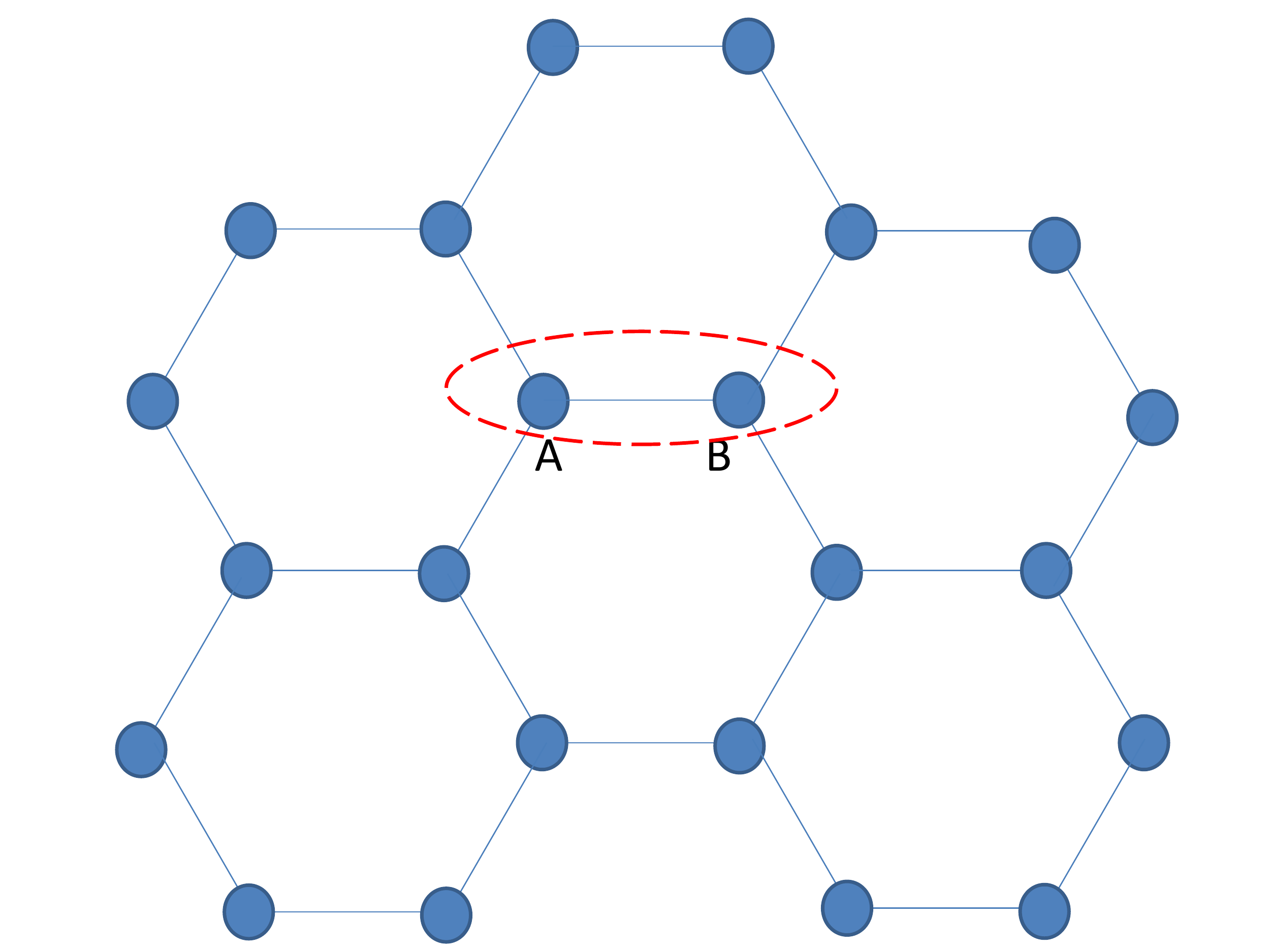}
\caption{Crystal structure of graphene.}
\label{figure0}
\end{figure}

The crystal structure of graphene is shown in Fig. \ref{figure0}. It is a Bravais lattice with a basis of two carbon atoms, which are labeled A and B. This crystal structure has many interesting consequences, among them is the focus of the rest of the paper, the Berry phase.

\begin{figure}[!ht]
\centering
\includegraphics[width=3in]{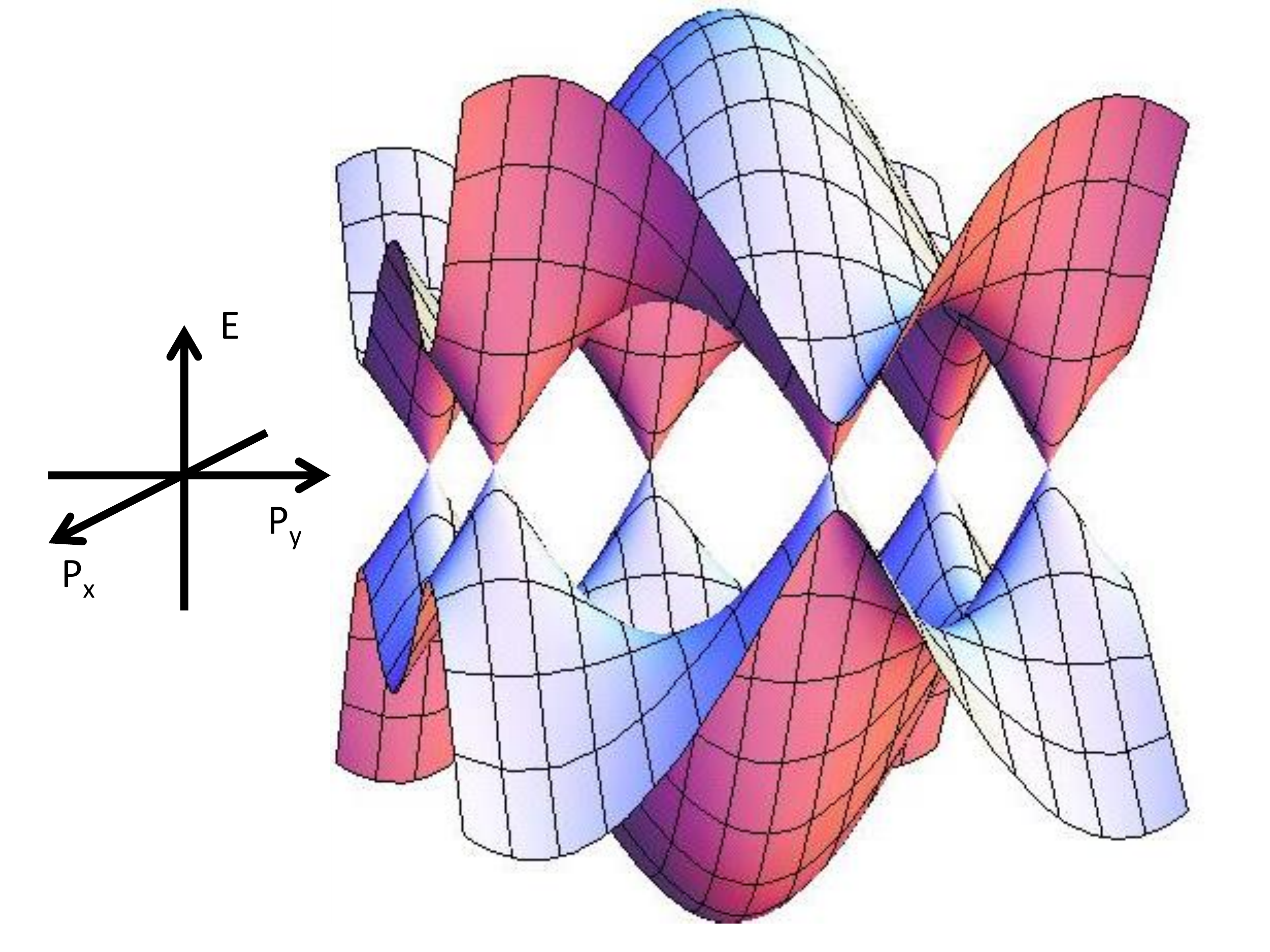}
\caption{Band structure of graphene.}
\label{figure1}
\end{figure}

\begin{figure}[!ht]
\centering
\includegraphics[width=3in]{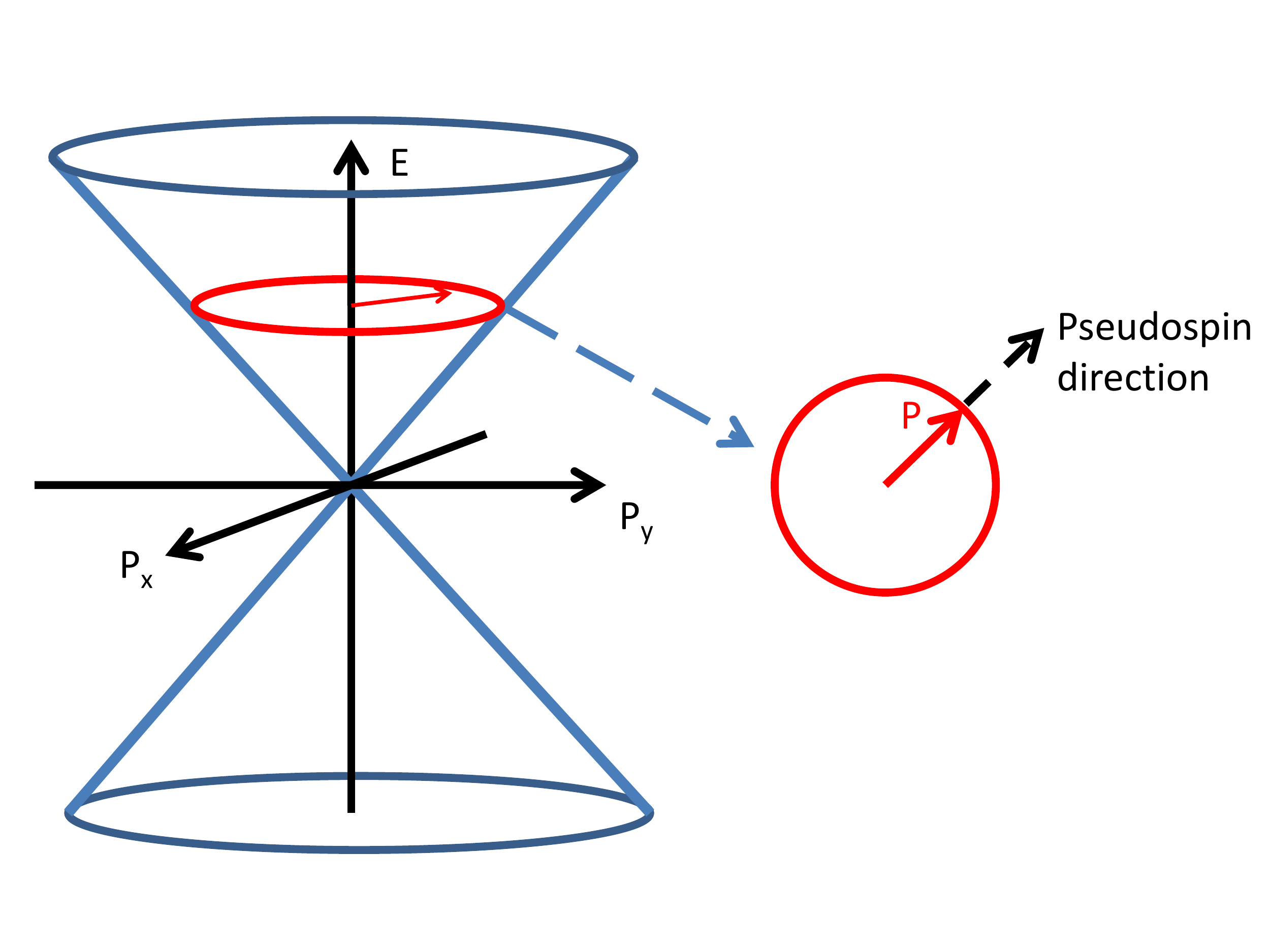}

\caption{Dirac cone and the pseudospin direction.}
\label{figure2}
\end{figure}

The tight binding calculation is often used to calculate the band structure of graphene. A very good review with all the details of the calculation can be found in Ref. \onlinecite{mcannpaper}. The calculated band structure is shown in Fig. \ref{figure1}. The conduction band and valence band touch each other at six points at the corners of the first Brillouin zone, which are the so called Dirac points. In most cases, we are only concerned with the low energy electronic states near the Dirac points, and the dispersion relation there is approximately linear \cite{mcannpaper}. The effective low energy Hamiltonian (near one of the Dirac points) can be written as \cite{mcannpaper}:
\begin{equation}
\label{hamiltonian}
H=
v\begin{pmatrix}
	 0 &  p_x-ip_y \\
	 p_x+ip_y & 0 \\
\end{pmatrix},
\end{equation}
where $v\approx 10^6 m/s$ is the Fermi velocity in graphene, and $p_x$ and $p_y$ are the crystal momenta measured from a Dirac point, as shown in Fig. \ref{figure2}. In this Hamiltonian, the diagonal elements represent the `onsite energy' of electrons residing completely on either A or B atoms (they are defined as zero energy in Eq. \ref{hamiltonian}), while the off-diagonal elements represent the `hopping' of an electron from A to B or vice versa. The eigenvalues $E$ and eigenstates $\psi$ of this Hamiltonian satisfy the equation $H\Psi=E\psi$ and can be solved to give
\begin{equation}
\label{eigen}
E_\pm = \pm vp, \quad \quad \psi=\frac{1}{\sqrt{2}}
\begin{pmatrix}
  1\\
  \pm e^{i\phi}\\
\end{pmatrix}
e^{i\textbf{p}\cdot\textbf{r}/\hbar},
\end{equation}
where $\pm$ correspond to conduction ($+$) and valence ($-$) bands, and $\phi$ is the angle between the crystal momentum $\textbf{p}$ and the x axis, i.e. $p_x=p\cos\phi$ and $p_y=p\sin\phi$. 

The two component vector part of the eigenstates is the so called pseudospin, since it resembles the two component real spin vector. The origin of this pseudospin comes from the two atoms in the basis of graphene lattice, and the two components represent the relative weight (and phase) of the wavefunctions residing on A atoms and B atoms. For example, $(1,0)^\text T$ means that the eigenstate consists of wavefunctions only from A atoms, while $(0,1)^\text T$ means the opposite. Due to the symmetry between A and B atoms, we can see from Eq. \ref{eigen} that they contribute equally to the total wavefunction, only differing by a phase factor.

\section{Berry phase}
\label{berry}

In the context of graphene, the Berry phase is the phase that an eigenstate in Eq. \ref{eigen} acquires after $\textbf{p}$ is forced to evolve for a full circle at constant energy around a Dirac point (For more information about the Berry phase in condensed matter physics, please refer to Ref. \onlinecite{niuqian}). Magnetic field is usually applied to make $\textbf{p}$ evolve in this way, so the Berry phase is closely related to magnetotransport measurements such as the quantum Hall effect, which will be discussed in the next section.

Due to the peculiar band structure of graphene, the Berry phase is $\pi$ instead of a trivial 0 or $2\pi$. To see where this phase comes from, we note that the eigenstates in Eq. \ref{eigen} are commonly rewritten via a gauge transformation as
\begin{equation}
\label{eigen2}
\psi=\frac{1}{\sqrt{2}}
\begin{pmatrix}
  e^{-i\phi/2}\\
  \pm e^{i\phi/2}\\
\end{pmatrix}
e^{i\textbf{p}\cdot\textbf{r}/\hbar}.
\end{equation}
It is tempting to think that as $\textbf{p}$ rotates in a full circle, $\phi$ changes by $2\pi$, so the pseudospin vector acquires a minus sign, which is equivalent to $e^{i\pi}$ and a Berry phase of $\pi$.  However this interpretation \cite{rev of modern physics} is not appropriate. We can see that if we adopt the representation of the eigenstate in Eq. \ref {eigen}, then changing $\phi$ by $2\pi$ leaves the state the same. Strictly speaking, the eigenstates in Eq. \ref{eigen2} are not a good choice because they are not single-valued functions of $\phi$. We will see later that this so-called $\pi$ ``phase shift'' coming from eigenstates being non-single-valued can add to the real $\pi$ Berry phase to give a wrong answer.

We can derive the $\pi$ phase appropriately in two ways. The first is to calculate the Berry phase from its definition; the second is to utilize the rotation property of a spin 1/2 angular momentum.

The Berry phase $\theta$ is defined as \cite{niuqian}
 \begin{equation}
\label{berry phase definition}
\theta =- i\oint_C\langle\psi(\mathbf{p}(t))|\frac{\partial}{\partial t}|\psi(\mathbf{p}(t))\rangle dt.
\end{equation}
The integral is carried out as the crystal momentum $\mathbf{p}$ is evolving in a closed circle C. Plugging in the eigenstate from Eq. \ref{eigen} we can get
 \begin{equation}
\label{berry phase pi}
\theta = -i\oint_C dt[\frac{1}{2}
\begin{pmatrix}
1, & e^{-i\phi}
\end{pmatrix}
\begin{pmatrix}
0 \\
i\frac{\partial {\phi}}{\partial t}e^{i\phi}
\end{pmatrix}
+i\frac{\partial \mathbf{p}}{\partial t}\cdot \mathbf{r}/\hbar+i\mathbf p\cdot \frac{\partial \mathbf{r}}{\partial t}/\hbar].
\end{equation}
Note that the second and third term in Eq. \ref{berry phase pi} are zero after integration over a closed loop, so we get $\theta =  \oint_C dt (\partial \phi/\partial t)/2 = \pi$.

If the eigenstate plugged into Eq. \ref{berry phase definition} was from Eq. \ref{eigen2} instead of Eq. \ref{eigen}, one could get a seemingly different answer, i.e. $\theta = 0$. However, this result is incorrect. As mentioned earlier, the eigenstates in Eq. \ref{eigen2} get a minus sign when we change $\phi$ from 0 to $2\pi$. This causes a $\pi$ ``phase shift'' that cancels the real Berry phase of $\pi$ to give the wrong answer. If the discontinuity of Eq. \ref{eigen2} is taken into account and the integration is done properly\cite{winding number}, we will get the same Berry phase of $\pi$.

Now let's view the Berry phase from a different perspective, i.e. the rotation of the pseudospin. First we can define the ``direction'' of the pseudospin\cite{mcannpaper}. To do that, we recognize that the Hamiltonian in Eq. \ref{hamiltonian} can be rewritten in a form with Pauli matrices as
\begin{equation}
\label{hamiltonian2}
H=v( p_x \sigma_x+p_y\sigma_y),
\end{equation}
where $\sigma_x$ and $\sigma_y$ are the Pauli matrices, which operate on the pseudospin part of the eigenstates. We can find the ``direction'' of the pseudospin by calculating the expectation values of $\sigma_x$, $\sigma_y$ and $\sigma_z$. It is easy to get that
\begin{equation}
\label{pseudospin direction}
\langle\psi|\sigma_x|\psi\rangle = \cos\phi,\quad \langle\psi|\sigma_y|\psi\rangle = \sin\phi \quad \text{and} \quad \langle\psi|\sigma_z|\psi\rangle = 0.
\end{equation}
We can see that the pseudospin is in the direction of $\mathbf p$, lying in the plane of x and y, as shown in Fig. \ref{figure2}. When $\mathbf p$ is forced to rotate in a full circle ($2\pi$), so does the pseudospin. We know from quantum mechanics\cite{sakurai} that rotating a spin 1/2 state along the z direction by an angle of $\phi$ can be done with the operator $e^{-i\phi S_z/\hbar}=e^{-i\phi\sigma_z/2}$. When $\phi$ is $2\pi$, the rotation operator brings a minus sign to the pseudospin vector, which can be viewed as multiplied by $e^{i\pi}$, hence the extra phase of $\pi$ is acquired.

\section{The connection between Berry phase and the unconventional quantum Hall effect}
\label{quantumhall}

The unconventional quantum Hall effect in graphene\cite{2005 geim paper, 2005 kim paper} is signatured by the shifted Hall plateaus. This shift comes from a zero energy Landau level for graphene. In conventional systems, such as free electron gas and GaAs/AlGaAs 2DEG, the Landau level energy can be written as $E_n = \hbar\omega_C(n+1/2)$, where n is a non-negative integer, $\omega_C$ is the cyclotron frequency of electrons. In those systems the lowest Landau level energy is $E_n = 1/2\hbar\omega_C$, the so called ``zero-point energy''. However, the Landau level energy in graphene is $E_n=v\sqrt{2neB\hbar}$, without a zero-point energy. This unconventional zero energy Landau level in graphene can be derived from the Hamiltonian in Eq. \ref{hamiltonian} with the applied magnetic field taken into account\cite{mcannpaper}. From a different point of view, it is also a direct consequence of the $\pi$ Berry phase discussed above. 

To see this, we need to invoke Onsager's semiclassical derivation of the Landau levels \cite{topological} for electrons in solids. When a magnetic field is applied, electrons are forced into circular orbits. Classically, any radius is allowed. But using the Bohr-Sommerfeld quantization rule, we need to satisfy condition: $\oint d\mathbf{r}\cdot\mathbf{p} /\hbar= 2\pi n$, where n is a non-negative integer. This condition is to ensure that the wavefunction is single valued.  For electrons in a magnetic field (vector potential $\mathbf A$), the momentum can be written as $\mathbf{p}=\hbar \mathbf{k}-e\mathbf{A}$. Plugging this into the Bohr-Sommerfeld quantization condition we get
\begin{equation}
\label{bohr}
(\hbar\oint d\mathbf r\cdot\mathbf{k} - e\oint d \mathbf {r}\cdot\mathbf {A})/\hbar = 2\pi n.
\end{equation}
Within the semiclassical model of Bloch electrons in magnetic field\cite{mermin}, we have $\hbar\dot{\mathbf k} = -e \dot{\mathbf r}\times {\mathbf B}$, which gives
\begin{equation}
\label{semiclassical}
\hbar\mathbf k = -e \mathbf r\times {\mathbf B}+constant.
\end{equation}
Plugging Eq. \ref{semiclassical} into the first term of Eq. \ref{bohr}, we get
\begin{equation}
\label{first term}
-e\mathbf B\cdot\oint d \mathbf r\times\mathbf r = 2e\Phi,
\end{equation}
where $\Phi$ is the magnetic flux of the orbit under the field $\mathbf B$. Using Stocks' theorem, the second term of Eq. \ref{bohr} can be written as $e\int d\mathbf S \cdot \nabla\times \mathbf A = e\Phi$. So Eq. \ref{bohr} can be written as
\begin{equation}
\label{final bohr}
\Phi = nh/e,
\end{equation}
where h/e is the flux quantum. This is the quantization condition in real space, which means that the magnetic flux threading the orbit of an electron should be quantized to an integer times the flux quantum. Since we have the relation between $\mathbf r$ and $\mathbf k$ in Eq. \ref{semiclassical}, we can translate this condition to momentum space as
\begin{equation}
\label{quantization in momentum space}
\pi k^2 = 2\pi n eB/\hbar.
\end{equation}
If we apply this condition to the free electron gas, we get the energy quantization rule
\begin{equation}
\label{landau level}
E_n=\hbar^2 k^2/2m=n\hbar eB/m=n\hbar \omega_C,
\end{equation}
where $\omega_C$ is the cyclotron frequency. Obviously the semiclassical method doesn't count for the `zero-point energy' of $1/2\hbar\omega_C$. We can rewrite the Bohr-Sommerfeld quantization condition of Eq. \ref{bohr} as
\begin{equation}
\label{bohr modified}
(\hbar\oint d\mathbf r\cdot\mathbf{k} - e\oint d \mathbf {r}\cdot\mathbf {A})/\hbar = 2\pi (n+1/2).
\end{equation}
to take this factor into account. This extra 1/2 is the so called Maslov index \cite{topological}, and it is a fully quantum mechanical effect which is not captured by the semiclassical model.

Now let's consider the quantization rule for graphene. As discussed in the previous section, when the momentum $\mathbf p$ of an electron is forced to evolve in a circle, in addition to the normal phase factors of $\oint d\mathbf r\cdot\mathbf{k}$ and $ e\oint d \mathbf {r}\cdot\mathbf {A}/\hbar$, it also acquires an extra Berry phase of $\pi$. So Eq. \ref{bohr modified} should be
\begin{equation}
\label{bohr modified again}
(\hbar\oint d\mathbf r\cdot\mathbf{k} - e\oint d \mathbf {r}\cdot\mathbf {A})/\hbar + \pi = 2\pi (n+1/2).
\end{equation}
Interestingly, the Berry phase cancels the Maslov index. So the final quantization rule in momentum space for graphene is the same as Eq. \ref{quantization in momentum space}. Since the dispersion relation for graphene is $E=v\hbar k$, the Landau level energy for graphene is
\begin{equation}
\label{landau level for graphene}
E_n=v\sqrt{2neB\hbar}.
\end{equation}
This result agrees with the full quantum mechanical calculation\cite{mcannpaper}. From this derivation, we can see that the zero-energy Landau level is indeed a consequence of the $\pi$ Berry phase.

\begin{acknowledgments}

The author gratefully acknowledges Matthew Yankowitz for his help with the paper.

\end{acknowledgments}

\end{document}